\documentclass[conference]{IEEEtran}
\IEEEoverridecommandlockouts
\usepackage{cite}
\usepackage{amsmath,amssymb,amsfonts}
\usepackage{algorithmic}
\usepackage{graphicx}
\usepackage{textcomp}
\usepackage{xcolor}
\def\BibTeX{{\rm B\kern-.05em{\sc i\kern-.025em b}\kern-.08em
    T\kern-.1667em\lower.7ex\hbox{E}\kern-.125emX}}
\ifCLASSOPTIONcompsoc
    \usepackage[caption=false, font=normalsize, labelfont=sf, textfont=sf]{subfig}
\else
\usepackage[caption=false, font=footnotesize]{subfig}
\fi
\usepackage[utf8]{inputenc}
\usepackage[T1]{fontenc}
\usepackage{bm}
\usepackage{url}
\DeclareMathOperator*{\maximize}{maximize}
\DeclareMathOperator*{\minimize}{minimize}
\begin{document}

\title{Uplink Energy Efficiency of Cell-Free \\ Massive MIMO With Transmit Power Control \\ in Measured Propagation Channels
\thanks{The work of T. Choi and A. F. Molisch is supported by KDDI Research, Inc. and the National Science Foundation (ECCS-1731694 and ECCS-1923601).}
}

\author{\IEEEauthorblockN{Thomas Choi\IEEEauthorrefmark{1}, Masaaki Ito\IEEEauthorrefmark{2}, Issei Kanno\IEEEauthorrefmark{2}, Takeo Oseki\IEEEauthorrefmark{2}, Kosuke Yamazaki\IEEEauthorrefmark{2}, and Andreas F. Molisch\IEEEauthorrefmark{1}}
\IEEEauthorblockA{\IEEEauthorrefmark{1}University of Southern California,
Los Angeles, CA, United States} 
\IEEEauthorblockA{\IEEEauthorrefmark{2}KDDI Research, Inc., Saitama, Japan} 
}

\maketitle

\begin{abstract}
Cell-free massive MIMO (CF-mMIMO) is expected to provide reliable wireless services for a large number of user equipments (UEs) using access points (APs) distributed across a wide area. When the UEs are battery-powered, uplink energy efficiency (EE) becomes an important performance metric for CF-mMIMO systems. Therefore, if the ``target'' spectral efficiency (SE) is met, it is important to optimize the uplink EE when setting the transmit powers of the UEs. Also, such transmit power control (TPC) method must be tested on channel data from real-world measurements to prove its effectiveness. In this paper, we compare three different TPC algorithms using zero-forcing reception by applying them to 3.5 GHz channel measurement data featuring $\sim$30,000 possible AP locations and 8 UE locations in a 200m×200m area. We show that the max-min EE algorithm is highly effective in improving the uplink EE at a target SE, especially if the number of single-antenna APs is large, circuit power consumption is low, and the maximum allowed transmit power of the UEs is high.
\end{abstract}

\begin{IEEEkeywords}
Cell-free (distributed) massive MIMO, energy efficiency, transmit power control, experimental evaluations, channel sounding, drone (UAV) channel sounder
\end{IEEEkeywords}

\section{Introduction}
\subsection{Motivation}
Cell-free massive MIMO (CF-mMIMO), which combines various wireless communication system concepts such as mMIMO, ultra-dense networks, and cooperative multi-point (CoMP), exploits a large number of access points (APs) distributed across a wide area to reliably serve a large number of user equipments (UEs) while eliminating the inter-cell interference conventional cellular systems suffer from \cite{demir2020foundations}. When the UEs transmit information to the APs during the uplink phase, the energy efficiency (EE) is an important performance metric, especially if the UEs are battery-powered and if high uplink spectral efficiency (SE) is not required. Hence, finding transmit power control (TPC) algorithms that can maximize the EE at a ``target'' SE is very important. 

It is also important to test such TPC algorithms on large amount of realistic channel data, as the wireless providers planning the deployment of CF-mMIMO systems require accurate and reliable statistics of the expected performance. Such real-world data can only be obtained from extensive channel measurement campaigns. However, such large measurement datasets for CF-mMIMO systems are scarce due to the complexity of setting up and operating a massive number of antennas simultaneously. To address this issue, we showed that a large amount of channel data for CF-mMIMO systems can be measured using a compact channel sounder with a drone acting as a virtual array \cite{choi2021using}. 

\subsection{Related Works}
There were numerous CF-mMIMO studies which tackled the problem of improving the EE. In order to save energy at the APs, efforts were made to maximize the \emph{total downlink EE}, with various precoding methods and operating frequencies \cite{nguyen2017energy, ngo2018on, alonzo2018energy, tran2019first, jin2020spectral, qiu2020downlink}. Other works analyzed the downlink EE while maximizing the \emph{minimum downlink SE} per UE, in the cases of hardware impairments \cite{zhang2018performance}, in comparison to cellular systems \cite{yang2018energy}, or in relation to security \cite{alageli2020optimal}. Likewise for uplink, the EE was analyzed to maximize the \emph{minimum uplink SE} among the UEs \cite{bai2019max-min, zhao2020efficient, zhang2020analysis, demir2021joint, yan2021scalable}. There were also efforts to optimize the power coefficients for both the uplink and downlink jointly while seeking a balance between the EE and SE \cite{nguyen2020on, wang2020wirelessly, zhang2021spectral}.

In \cite{ito2021}, we suggested the max-min EE method, which optimizes the power coefficients to maximize the \emph{minimum uplink EE} over all UEs. This algorithm achieved improved EE for UEs with the lowest EE. However, all the works mentioned above including our previous work were based on simulated channel data achieved from statistical channel models. While \cite{choi2021using} observed the uplink SE of CF-mMIMO systems using channel measurement data, it lacked analysis of EE and impact of TPC.

\subsection{Contributions}
In this paper, we apply three different TPC algorithms (max-power, max-min SE, and max-min EE) to a large number of measured channel data to analyze the trade-offs between the EE and SE for CF-mMIMO systems with varying numbers of single-antenna APs. The amount of data is immense, featuring $\sim$30,000 possible AP locations and 8 UE locations in a 200m$\times$200m area at a university campus, providing statistical confidence of the evaluated performances in a realistic deployment setting. We show that the max-min EE is very effective in improving the uplink EE at a target SE, especially if the number of single-antenna APs is large, circuit power consumption is low, and the maximum allowed transmit power of the UEs is high.

\section{System Model} \label{system}
\begin{figure}[!t]
\centering
\includegraphics[width=\linewidth]{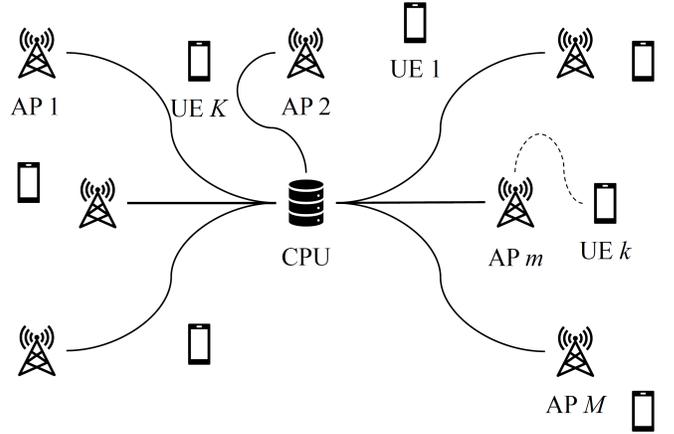}
\caption{Model of a cell-free massive MIMO system, with $M$ single-antenna APs and $K$ single-antenna UEs}
\label{fig:cfmmimo}
\end{figure}

We consider a CF-mMIMO system, with $M$ single-antenna APs deployed in a selected area of service (see \figurename~\ref{fig:cfmmimo}).\footnote{We assume unlimited backhaul; limited backhaul is a separate topic beyond the scope of this work.} The total number of single-antenna UEs served within the same channel resource block is $K$. We consider frequency-flat fading, such that the channel coefficient between AP $m$ and UE $k$ is:

\begin{IEEEeqnarray}{rCl}
h_{m,k}&=&\sqrt{\beta_{m,k}}p_{m,k},%
\end{IEEEeqnarray}
where $\beta_{m,k}$ and $p_{m,k}$ are large- and small-scale fading of the corresponding links, respectively. 

\subsection{Uplink System Model}
The received signal at antenna $m$ on the network side is:

\begin{IEEEeqnarray}{rCl}
y_m&=&\sqrt{\rho}\sum_{k=1}^K h_{m,k}\sqrt{q_k}s_k+z_m,%
\end{IEEEeqnarray}
where $s_k$ is a transmitted symbol of UE $k$ normalized to unit average power, $0 \leq q_k \leq 1$ is the transmit power coefficient, $z_m\sim\mathcal{N}_{\mathbb{C}}\mathopen{}\left(0,1\right)\mathclose{}$ is the normalized noise, and $\rho$ is the transmit SNR, i.e., the ratio of the maximum transmitted signal power to the noise power.

\subsection{Channel Estimation}
For channel estimation, $\tau^{\mathopen{}\left(\text{p}\right)\mathclose{}}$-length pilot resources from each UE are consumed within the coherence interval.\footnote{We assume all UEs use orthogonal pilot sequences ($\tau^{\mathrm{(p)}}=K$).} Let $\sqrt{\tau^{\mathopen{}\left(\text{p}\right)\mathclose{}}}\bm{\varphi}_k$ be the $\tau^{\mathopen{}\left(\text{p}\right)\mathclose{}}$-dimensional pilot sequence vector of UE $k$, where $\mathopen{}\left\|\bm{\varphi}_k\right\|\mathclose{}^2=1$, and the corresponding received signal vector can be written as:

\begin{IEEEeqnarray}{rCl}
\bm{y}_m^{\mathopen{}\left(\text{p}\right)\mathclose{}}&=&\sqrt{\rho^{\mathopen{}\left(\text{p}\right)\mathclose{}}\tau^{\mathopen{}\left(\text{p}\right)\mathclose{}}}\sum_{k=1}^Kh_{m,k}\bm{\varphi}_k+\bm{z}_m^{\mathopen{}\left(\text{p}\right)\mathclose{}}.%
\end{IEEEeqnarray}

The MMSE channel estimate can then be written as \cite{ohurley2020comparison}:

\begin{IEEEeqnarray}{rCl}
\Hat{h}_{m,k}&=&\frac{\sqrt{\rho^{\mathopen{}\left(\text{p}\right)\mathclose{}}\tau^{\mathopen{}\left(\text{p}\right)\mathclose{}}}\beta_{m,k}}{\rho^{\mathopen{}\left(\text{p}\right)\mathclose{}}\tau^{\mathopen{}\left(\text{p}\right)\mathclose{}}\sum_{k'=1}^K\beta_{m,k'}\mathopen{}\left|\bm{\varphi}_k^\text{H}\bm{\varphi}_{k'}\right|\mathclose{}^2+1}\bm{\varphi}_k^{\text{H}}\bm{y}_m^{\mathopen{}\left(\text{p}\right)\mathclose{}}.%
\end{IEEEeqnarray}

\section{Performance Metrics}
To analyze the performances of different TPC algorithms in Sec. \ref{ener_eff}, we evaluate the SE and EE. We assume zero-forcing (ZF) reception on the network side, due to its reduced inversion complexity than the linear minimum mean square error (LMMSE) receiver when $M>>K$ \cite{demir2020foundations}.

For ZF reception, the central processing unit (CPU) collects the received signals from all antennas, vectorized as:

\begin{IEEEeqnarray}{rCl}
\bm{y}=\sqrt{\rho}\bm{H}\bm{Q}^{1/2}\bm{s}+\bm{z},%
\end{IEEEeqnarray}
where the dimension of $\bm{H}$ is $M\times K$ ($\bm{H}(m,k) = h_{m,k}$) and the dimension of the diagonal matrix $\bm{Q}$ is $K\times K$ ($\bm{Q}(k,k) = q_{k}$). The $K\times M$ ZF weight matrix is formed as:

\begin{IEEEeqnarray}{rCl}
\bm{W}=&\mathopen{}\left(\Hat{\bm{H}}^{\text{H}}\Hat{\bm{H}}\right)\mathclose{}^{-1}\Hat{\bm{H}}^{\text{H}},%
\end{IEEEeqnarray}
where $\Hat{\bm{H}}$ is the estimate of $\bm{H}$. The channel estimation error matrix is defined as $\Tilde{\bm{H}}=\bm{H}-\Hat{\bm{H}}$, i.e., $h_{m,k}=\Hat{h}_{m,k}+\tilde{h}_{m,k}$.

\subsection{Spectral Efficiency}
The SE of UE $k$ with ZF is formulated as:

\begin{IEEEeqnarray}{rCl}
S_k
&=&\log_2\mathopen{}\left(1+\frac{\rho q_k}{\rho\sum_{k'\ne k}^Kq_{k'}\mathopen{}\left|\bm{w}_k^\text{H}\Tilde{\bm{h}}_{k'}\right|\mathclose{}^2+\mathopen{}\left\|\bm{w}_k\right\|\mathclose{}^2}\right)\mathclose{},\label{eq:SE}%
\end{IEEEeqnarray}
where the vector $\bm{w}_k = \bm{W}(k,:)^\text{T}$ with the dimension $M\times 1$.

\subsection{Energy Efficiency}
Based on~\cite{bashar2019energy}, the power consumption of UE $k$ is:

\begin{IEEEeqnarray}{rCl}
P_{k}&=&\Bar{P}q_k+P_{\text{U}},\label{eq:ee_rev}%
\end{IEEEeqnarray}
where $\Bar{P}$ is the maximum allowed transmit power and $P_{\text{U}}$ is the required power to run circuit components at each UE. 

The EE of UE $k$ is defined as:

\begin{IEEEeqnarray}{rCl}
E_k&=&\frac{\text{Bandwidth}\cdot S_k}{P_{k}}.\label{eq:ee_ue}%
\end{IEEEeqnarray}

\section{Transmit Power Control Algorithms} \label{ener_eff}
TPC is needed to improve the EE of CF-mMIMO systems. In this work, we consider the uplink case, prioritizing the EE of the UEs. We introduce three different types of TPC algorithms: max-power, max-min SE, and max-min EE.

\subsection{Max-Power Method}
Max-power is the most simplistic method: each UE transmits with the maximum allowed power ($q_k=1$). It is not strictly a TPC method, but we use it as the baseline to be compared with other TPC algorithms, as it is expected to produce highest SE cancelling interference from other UEs under ZF reception and lowest EE.

\subsection{Max-Min Spectral Efficiency Method}
Max-min SE is one of the most commonly used TPC methods in CF-mMIMO literature, and aims to maximize the minimum SE among all UEs.

The optimization problem is written as:

\begin{IEEEeqnarray}{ll}
\maximize_{\mathopen{}\left\{q_k\right\}\mathclose{}}&\min_{k=1,\dots,K}S_k\label{eq:prob}\\
\text{subject to }&0\le q_k\le1,k=1,\dots,K.\IEEEnonumber%
\end{IEEEeqnarray}

Since the SE is a logarithmic function increasing monotonically with the signal-to-interference-plus-noise ratio (SINR), the problem \eqref{eq:prob} can be reformulated as:

\begin{IEEEeqnarray}{ll}
\maximize_{\mathopen{}\left\{q_k\right\}\mathclose{},t}&t\label{eq:prob2}\\
\text{subject to }&t\le\text{SINR}_k,k=1,\dots,K\IEEEnonumber\\
&0\le q_k\le1,k=1,\dots,K.\IEEEnonumber%
\end{IEEEeqnarray}

As proved in~\cite{bashar2019uplink}, the problem \eqref{eq:prob2} is formulated into a standard geometric programming problem, and can be solved by a software solver such as CVX for MATLAB~\cite{grant2020cvx,grant2008graph}.

\subsection{Max-Min Energy Efficiency Method}

To improve the EE at a given SE, \cite{ito2021} proposed the max-min EE TPC method. Similar to the max-min SE, the optimization problem of the max-min EE method is written as:

\begin{IEEEeqnarray}{ll}
\maximize_{\mathopen{}\left\{q_k\right\}\mathclose{}}&\min_{k=1,\dots,K}w_k^{\mathopen{}\left(\text{b}\right)\mathclose{}}E_k\label{eq:mm_ee_prob}\\
\text{subject to }&S_k\ge S_k^{\mathopen{}\left(\text{r}\right)\mathclose{}},k=1,\dots,K\IEEEnonumber\\
&0\le q_k\le 1,k=1,\dots,K,\IEEEnonumber%
\end{IEEEeqnarray}
where $w_k^{\mathopen{}\left(\text{b}\right)\mathclose{}}$ is the weight for each UE that can be chosen arbitrarily (e.g., proportional to the remaining battery charge), $S_k^{\mathopen{}\left(\text{r}\right)\mathclose{}}$ is the required minimum (target) SE for UE $k$ to ensure a certain level of quality of service. For this paper, we assume $w_k^{\mathopen{}\left(\text{b}\right)\mathclose{}}$ and $S_k^{\mathopen{}\left(\text{r}\right)\mathclose{}}$ are the same for all UEs, and denote them as 1 and $S^{\mathopen{}\left(\text{r}\right)\mathclose{}}$. Using the definition of EE, the problem \eqref{eq:mm_ee_prob} can be reformulated as:

\begin{IEEEeqnarray}{ll}
\maximize_{\mathopen{}\left\{q_k\right\}\mathclose{}}&\min_{k=1,\dots,K}\frac{\text{Bandwidth}\cdot S_k}{\Bar{P}q_k+P_{\text{U}}}\label{eq:mm_ee_prob_frac}\\
\text{subject to }&S_k\ge S^{\mathopen{}\left(\text{r}\right)\mathclose{}},k=1,\dots,K\IEEEnonumber\\
&0\le q_k\le 1,k=1,\dots,K.\IEEEnonumber%
\end{IEEEeqnarray}

To make the problem easier to handle, replace $q_k$ in the denominator with an auxiliary variable $\nu$:

\begin{IEEEeqnarray}{ll}
\maximize_{\mathopen{}\left\{q_k\right\}\mathclose{},\nu}&\min_{k=1,\dots,K}\frac{\text{Bandwidth}\cdot S_k}{\Bar{P}\nu+P_{\text{U}}}\label{eq:mm_ee_prob2}\\
\text{subject to }&S_k\ge S^{\mathopen{}\left(\text{r}\right)\mathclose{}},k=1,\dots,K\IEEEnonumber\\
&0\le q_k\le 1,k=1,\dots,K\IEEEnonumber\\
&q_k\le\nu,k=1,\dots,K\IEEEnonumber\\
&\nu^*\le\nu\le 1,\IEEEnonumber%
\end{IEEEeqnarray}
where $\nu^*$ is the slack variable and given as the maximum $q_k$ that achieves $S^{\mathrm{(r)}}$ obtained by solving the following optimization problem:

\begin{IEEEeqnarray}{ll}
\minimize_{\mathopen{}\left\{q_k\right\}\mathclose{}}&\max_{k=1,\dots,K}q_k\\
\text{subject to }&S_k\ge S^{\mathopen{}\left(\text{r}\right)\mathclose{}},k=1,\dots,K\IEEEnonumber\\
&0\le q_k\le 1,k=1,\dots,K,\IEEEnonumber%
\end{IEEEeqnarray}
which is explained further in \cite{ito2021}. Then, the problem \eqref{eq:mm_ee_prob2} can be solved as:
\begin{enumerate}
\item Finding the optimal value of $\nu$ to maximize the minimum EE using a linear search algorithm.
\item Optimizing $q_k$ to minimize the maximum transmit power when $S^{(\mathrm{r})}$ is reached.
\end{enumerate}

When solving the EE-maximization problem, $\nu$ is always the maximum value of $q_k$ while $\nu^*$ is the maximum value of $q_k$ to achieve the required SE. Therefore, the actual EE, which is calculated by using the optimized $q_k$, becomes higher than is calculated within the optimization problem because the actual denominator of EE for those UEs also becomes smaller ($\Bar{P}q_k+P_{\text{U}}\leq\Bar{P}\nu+P_{\text{U}}$).

\section{Channel Measurement} \label{chan_meas}
\subsection{Channel Sounder}
\begin{figure}[t!]
     \centering
     \includegraphics[width=\linewidth]{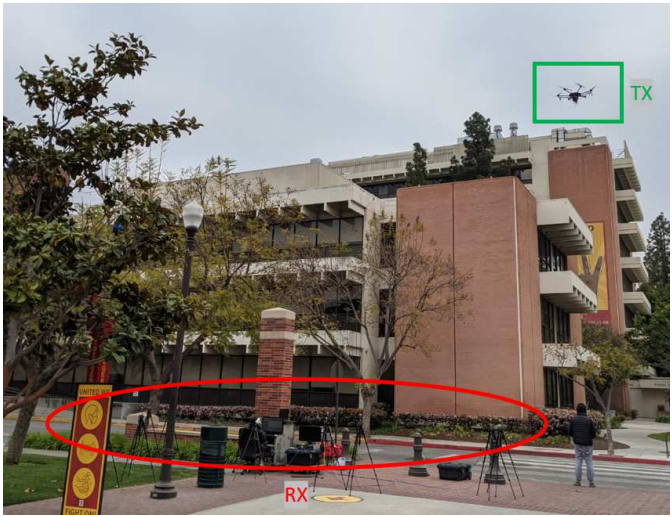}
     \caption{Channel sounder setup: a single TX antenna on a drone passes through a trajectory containing all possible AP locations while 8 separated RX antennas on the ground are positioned at 8 different UE locations}
     \label{fig:sndr}
\end{figure}

We acquired our channel data with a sounder that includes a transmitter (TX) on a drone \cite{ponce2021air} and a receiver (RX) on the ground (Fig. \ref{fig:sndr}). The TX acts as a \emph{virtual array} with a single antenna being moved by the drone to different locations, while the RX contains a switched array of eight physically separated antennas:\footnote{The separations of the eight antennas per measurement are limited by the 25 ft cables connecting the antennas to the switch.} the drone constantly transmits a 46 MHz OFDM-like sounding waveform with 2301 subcarriers (20 kHz subcarrier spacing) at 3.5 GHz as it moves at 1 m/s speed, passing through all locations where the APs may potentially be installed, while the RX constantly captures the channel data between the TX antenna and 8 RX antennas at 8 different UE locations every 50 ms through switching.\footnote{Note while the measurements were conducted from the AP side to the UE side, we can still analyze the uplink scenarios due to channel reciprocity. Responses of measurement equipment are removed through calibration \cite{ponce2021air}.} 
Therefore, 8$\times$2301 channel matrix may be captured at every 5 cm of drone movement. The characteristics of the drone sounder and the channel measurement principle for CF-mMIMO systems are further discussed in \cite{choi2021using, ponce2021air}.

\subsection{Channel Measurement Setting} \label{setting}
\begin{figure}[t!]
     \centering
     \includegraphics[width=\linewidth]{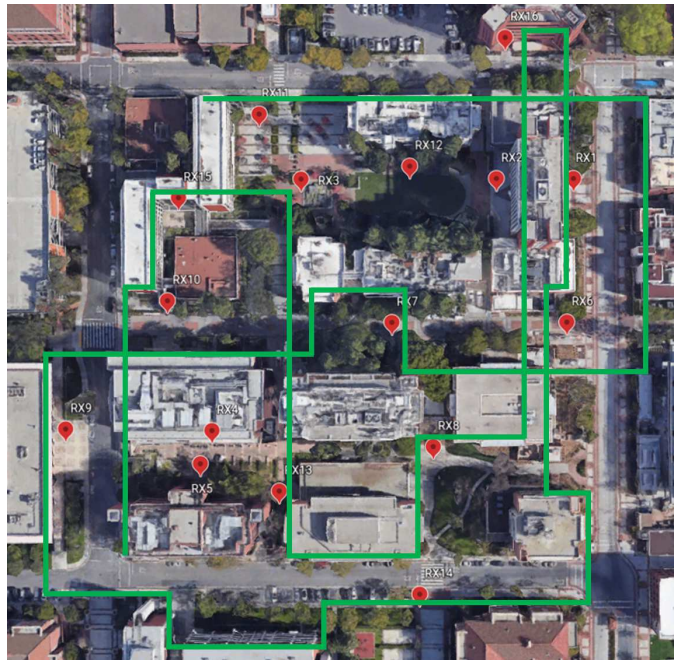}
     \caption{Overall channel measurement setting: the trajectory of the TX around 200m$\times$200m area at USC is shown by the green line, and the same trajectory was repeated 16 times for 16 different RX sites marked in red with 8 UE antennas per RX site, resulting in channel data for $\sim$30,000 possible AP and 128 UE locations - this paper considers 8 UE locations from RX3 site}
     \label{env}
\end{figure}

The channel measurements were conducted at the southwest side of the University of Southern California (USC) University Park Campus, as shown in Fig. \ref{env}. For each measurement, a RX site was selected within a 200m$\times$200m area to place the 8 UE antennas at 1.5m height, while the drone flew at 25 to 45m height (varying across the trajectory depending on the height of the building the drone flies over), autonomously through a software app. Since the drone moved at 1m/s through $\sim$1500m TX trajectory and there were in total 16 RX sites capturing channel matrix every 0.05s, channel data between $\sim$30,000 possible AP and 128 UE locations were attained. In this paper, 8 UEs in RX3 are chosen for evaluations in Sec. \ref{perf}.

\subsection{Applying Measurement Data to Analysis}
In our analysis, we consider the channel coefficient at each frequency index as a particular instance of a 3.5 GHz channel rather than considering each frequency index separately to account for frequency selectivity. Relating to the system model, $\bm{H}^{(i)}$ is the channel matrix between $M$ selected AP locations from the TX trajectory and $K$ selected UE locations from the RX sites at an instance $i$, and $\beta_{m,k} = \frac{1}{F}\sum_{i=1}^{F}(h_{m,k}^{(i)})^2$ is the average large-scale fading with $F$ the total number of instances corresponding to the number of frequency points from the measurement ($F=2301$). Calibration and time invariance of the sounder characteristics over the duration of the measurements were tested, and some frequency points (less than 10\% of the total acquired data) exhibiting calibration errors were discarded. 

\section{Performance Evaluations} \label{perf}
Performances of the TPC algorithms in Sec. \ref{ener_eff} are evaluated and compared by applying them to the channel data obtained from the measurement campaigns described in Sec. \ref{chan_meas}, using various setup parameters. We fix 20 MHz of bandwidth, 290K temperature, and 9 dB noise figure for all simulations. 

\subsection{Comparing Different Energy Efficiency Algorithms}
First, we compare different TPC algorithms and evaluate their trade-offs. For this comparison, 512 single-antenna APs ($M=512$) are chosen randomly from $\sim$30,000 possible locations, and 8 UEs from RX3 are considered. For the transmit power ($\bar{P}$) and the circuit power ($P_\text{U}$), 0.2W and 0.1W are assumed. 

\begin{figure}[!t]
    \centering
    \subfloat{\includegraphics[width=\linewidth]{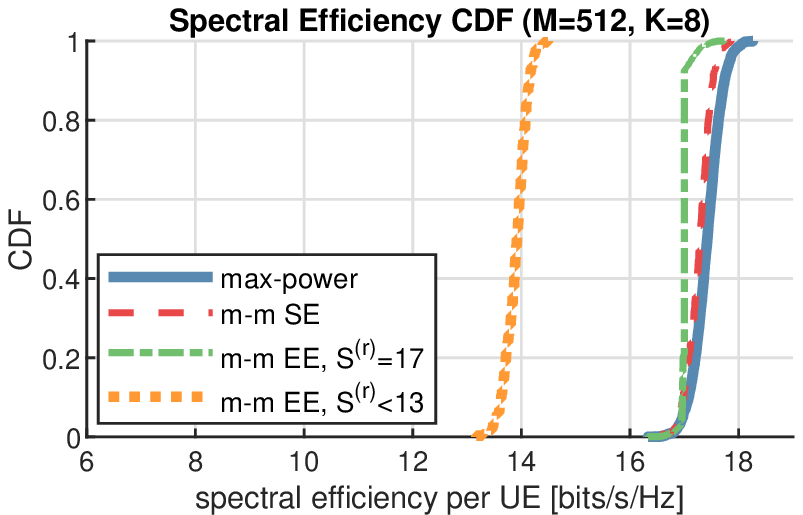}%
    \label{fig:SE}}
    \newline
    \subfloat{\includegraphics[width=\linewidth]{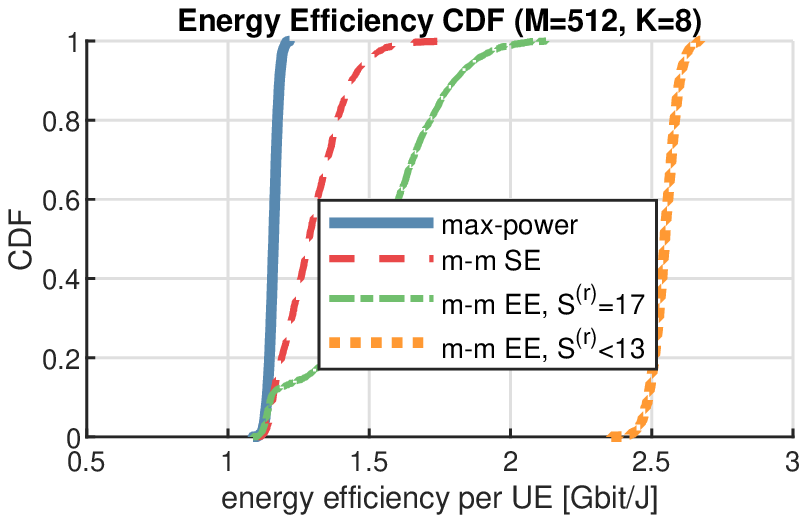}%
    \label{fig:EE}}
    
    \caption{CDF of spectral and energy efficiency for three different TPC algorithms, when M=512 and K=8}
    \label{fig:SE_EE}
\end{figure}

Fig. \ref{fig:SE_EE} shows how the SE generally ranks in the order of max-power, max-min SE, and max-min EE algorithm, while the EE ranks conversely. There is a clear trade-off between the SE and EE. It must be noted that the performance of the max-min EE algorithm can differ largely depending on $S^{(\mathrm{r})}$. For example, if we compare the max-min EE plots with two different values of $S^{(\mathrm{r})}$ (17 and $<$13)\footnote{Results for the max-min EE are the same if the $S^{(\mathrm{r})}$ is less than 13 bits/s/Hz; the maximum EE values are reached for those values of $S^{(\mathrm{r})}$.} on Fig. \ref{fig:SE_EE}, the median of the SE is greater for $S^{(\mathrm{r})}=$ 17 bits/s/Hz than $S^{(\mathrm{r})}<$13 bits/s/Hz by about 3 bits/s/Hz, while the median of the the EE is 1 Gbit/J less. If we increase the $S^{(\mathrm{r})}$ even further than 17 bits/s/Hz, both the SE and EE plots for the max-min EE approach the plots of the max-power algorithm. Under ZF reception, the max-min EE hence is a very flexible algorithm where the $S^{(\mathrm{r})}$ acts as an adjustable parameter modifying the system performances depending on the SE or EE requirements. The max-min SE meanwhile provides similar performance as the max-power for SE, while for EE, the max-min SE is better than the max-power by 0.13 Gbit/J at the median. 

\begin{figure}[!t]
    \centering
    \subfloat{\includegraphics[width=\linewidth]{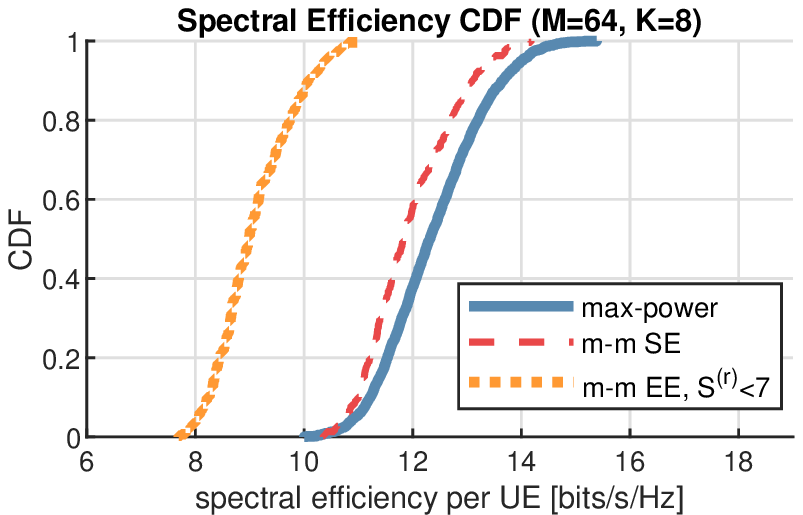}%
    \label{fig:SE_n}}
    \newline
    \subfloat{\includegraphics[width=\linewidth]{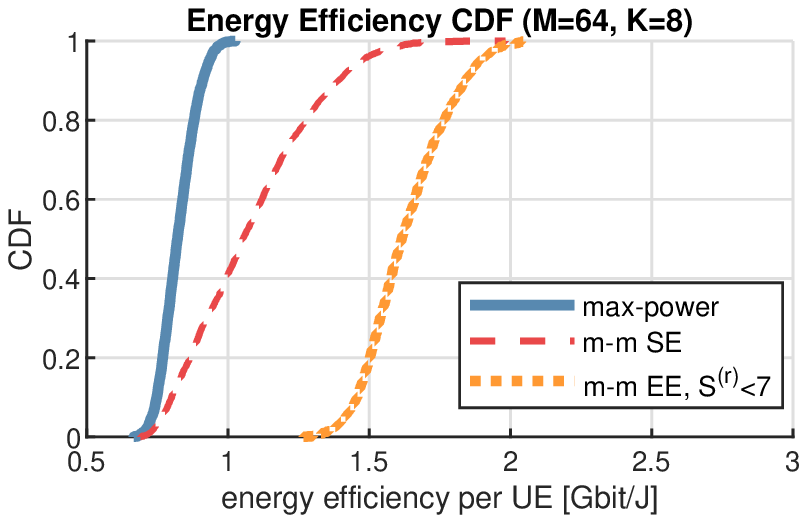}%
    \label{fig:EE_n}}
    
    \caption{CDF of spectral and energy efficiency for three different TPC algorithms, when M=64 and K=8}
    \label{fig:SE_EE_n}
\end{figure}

\begin{figure}[!t]
    \centering
    \subfloat{\includegraphics[width=\linewidth]{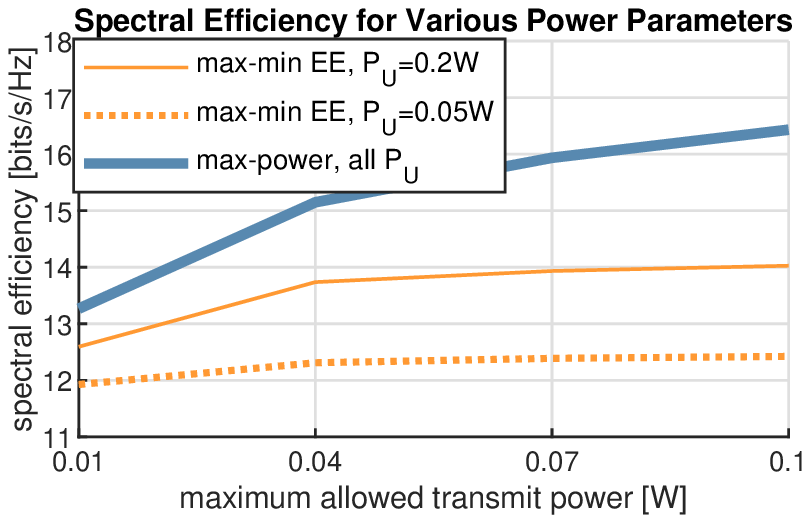}%
    \label{fig:SE_p}}
    \newline
    \subfloat{\includegraphics[width=\linewidth]{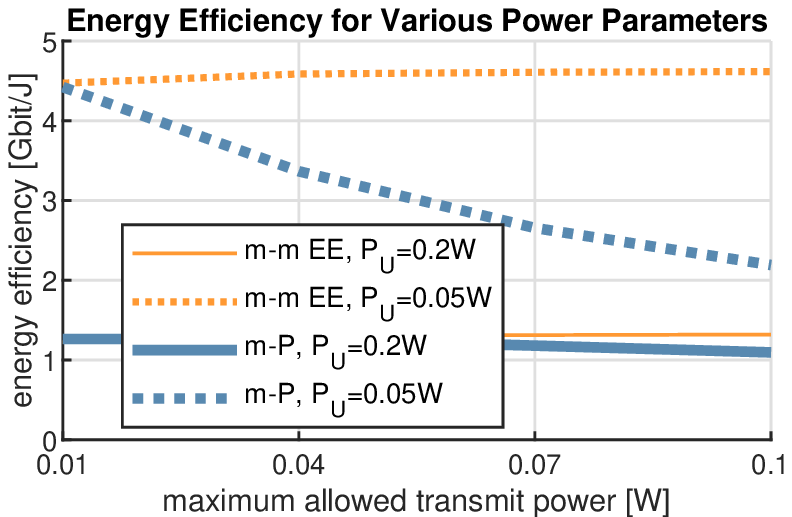}%
    \label{fig:EE_p}}
    
    \caption{Spectral and energy efficiency for various values of power parameters with $M=512$ and $K=8$}
    \label{fig:SE_EE_p}
\end{figure}

\subsection{Comparing Different Number of APs}
Now we reduce the number of single-antenna APs to 64 ($M=64$) while keeping the number of single-antenna UEs to 8 ($K=8$). For the max-min EE algorithm, we only observe the case when the EE is at the maximum (low $S^{(\mathrm{r})}$) since the max-min EE curve can lie anywhere between the max-power case and the maximum EE case depending on the $S^{(\mathrm{r})}$.

Comparing Fig. \ref{fig:SE_EE} and Fig. \ref{fig:SE_EE_n}, the plots are steeper for all algorithms for both the SE and EE when there is a larger number of APs, implying the uniform quality of service. Also, the plots for the max-min SE and max-min EE are closer to one another for the EE when there is less number of APs, indicating that the max-min EE can be more advantageous in terms of the EE when the number of single-antenna APs increase. Meanwhile, comparing $M=512$ and $M=64$, the distances at the median between the SE for the max-power and max-min EE are about 3.5 and 3.3 bits/s/Hz, while for the EE, there are 1.4 and 0.8 Gbit/J differences.

\subsection{Comparing Different Power Settings}
Lastly, we compare the performances at different power settings. Two power parameters which contribute to the EE are the maximum allowed transmit power ($\bar{P}$) and the power to run circuit components ($P_\text{U}$). We keep the number of APs and UEs to 512 and 8 respectively. For the max-min EE algorithm, we only select the cases where the EE is maximized (low $S^{(\mathrm{r})}$). We select the median values among all cases per power setting scenario. 

Fig. \ref{fig:SE_EE_p} shows the median SE and EE for various values of power parameters. Note that the SE doesn't change with $P_\text{U}$ for the max-power algorithm, since the SE depends on $\bar{P}q_k$, and $q_k$ is always assumed to be 1 regardless of $P_\text{U}$. Unlike the max-power algorithm, max-min EE algorithm shows that the SE does not increase much with $\bar{P}$. This is because even when $\bar{P}$ is increased, the UE does not transmit with maximum power to conserve energy ($q_k$ lies between 0 and 1).

For both the max-power and max-min EE algorithms, the EE increases when $P_\text{U}$ decreases. While the EE decreases with $\bar{P}$ for the max-power algorithm, the max-min EE algorithm has a constant EE regardless of $\bar{P}$. This again indicates that for the max-min EE, the UE will not transmit at a higher power even if the $\bar{P}$ increases since the EE is the priority. The max-min EE is also more advantageous over the max-power in terms of EE when the $\bar{P}$ increases and when $P_\text{U}$ is low, but the differences for the SE also grow. Again, depending on the $S^{(\mathrm{r})}$, the max-min EE curve can lie anywhere between the two curves at each power setting. Hence, for the max-min EE algorithm, the max-min EE curves maximizing EE acts as a lower bound for the SE and an upper bound for the EE, while the max-power acts as the upper bound for the EE and a lower bound for EE, under ZF reception.

\section{Conclusion and Future Work}
The max-min EE is a flexible TPC algorithm which can either provide the maximum EE by trading off the SE or provide the same performance as the max-power algorithm maximizing the SE with the least EE depending on the target SE ($S^{(\mathrm{r})}$). Max-min EE is especially effective than the max-power and max-min SE TPC algorithms in terms of the EE with higher number of single-antenna APs, lower circuit consumption power, and higher maximum allowed transmit power. Our future work will include comparing all measured channel data with statistical channel models to model CF-mMIMO channels as well as system analysis using varying antennas per AP, number of APs, number of UEs, and combining methods.

\bibliography{reference.bib} 
\bibliographystyle{IEEEtran}

\end{document}